\documentclass[twocolumn]{webofc}
\usepackage[varg]{txfonts}   % Web of Conferences font
\usepackage{amsmath}
\begin{document}
\title{Does Compound Nucleus remember its Isospin- An Evidence from the Fission Widths}

\author{\firstname{Swati} \lastname{Garg}\inst{1}\fnsep\thanks{\email{swat90.garg@gmail.com}} \and
        \firstname{Ashok Kumar} \lastname{Jain}\inst{1}\fnsep}%\thanks{\email{swat90.garg@gmail.com}}}% \and
%        \firstname{Third author} \lastname{Third author}\inst{3}\fnsep\thanks{\email{Mail address for last
%             author if necessary}}
%        % etc.
%}

\institute{Department of Physics, Indian Institute of Technology Roorkee, Roorkee-247667, India}

\abstract{%
  We present an evidence of isospin effects in nuclear fission by comparing the fission widths for reactions involving different isospin states of the same compound nucleus (CN). Yadrovsky~\cite{yadrovsky} suggested this possibility in 1975. Yadrovsky obtained the fission widths for two reaction data sets, namely $^{206}$Pb($\alpha$,f) and $^{209}$Bi(p,f), both leading to same CN, and concluded that “a nucleus remembers the isospin value of the nuclear states leading to fission”. We obtain the fission decay widths for both the $T_0+{\dfrac{1}{2}}$ and $T_0-{\dfrac{1}{2}}$ states of CN by using two appropriate reaction data sets. We then compare the fission widths for the two isospin states of CN. More specifically, we have chosen the combination of $^{206}$Pb($\alpha$,f) and $^{209}$Bi(p,f) same as presented in Yadrovsky's paper~\cite{yadrovsky} in this study. A significant difference between the ratios of fission decay widths to total decay widths for different isospin values suggests that isospin plays an important role in fission.
}
\maketitle
\section{Introduction}
\label{intro}
Isospin is a very simple but useful concept in nuclear physics. As the name suggests, it behaves in the same manner as spin. It is important from both structural and reaction point of view~\cite{robson}. In the early days since its discovery, it was considered to be useful only for light nuclei because it was considered to be pure for these nuclei. As we go towards heavy nuclei, isospin mixing starts increasing and it no longer remains a good quantum number. However, in 1962, Lane and Soper~\cite{lane} argued on theoretical basis that isospin impurity decreases as we go towards neutron-rich nuclei. The extra neutrons dilute the isospin impurity and isospin again becomes a nearly good quantum number in neutron-rich nuclei. We have used this concept of isospin conservation in reproducing the yields of neutron-rich fragments~\cite{jain,swati}. 

In 1975, Yadrovsky~\cite{yadrovsky} calculated the fission branching ratios for two possible isospin states $T_0+{\dfrac{1}{2}}$ and $T_0-{\dfrac{1}{2}}$ of CN for a combination of $^{206}$Pb($\alpha$,f) and $^{209}$Bi(p,f) reactions, both leading to same CN. It was concluded that the CN remembers isospin of initial nuclei. However, the author in Ref.~\cite{yadrovsky} was not certain whether this could lead to isospin conservation in fission or not. In the present work, we try to test Yadrovsky idea. We also calculate the same quantity for the combination taken by Yadrovsky but with a different set of experimental data. Our calculations provide a strong support to the concept that isospin is actively involved in fission.

\section{Formalism}
\label{sec-1}
We, first, consider a combination of proton and alpha induced reactions giving same CN. Suppose a proton with $T=\dfrac{1}{2}$, $T_3=\dfrac{-1}{2}$ is incident on a target with $T=T_3=T_0$ forming a CN. From isospin algebra, the possible values for isospin of CN are, $T_{CN}=T_0+\dfrac{1}{2}, T_0-\dfrac{1}{2}$ and $T_3= T_0-\dfrac{1}{2}$. Similarly, for $\alpha$ induced reaction on a target with $T'=T'_{3}=T_0-\dfrac{1}{2}$, there is only one possible isospin state of CN, $T_{CN}=T_{3_{CN}}=T_0-\dfrac{1}{2}$ since $\alpha$ particle has isospin $T=T_3=0$. The formalism, we are going to use is summarized by Yadrovsky in his paper~\cite{yadrovsky}. Yadrovsky deduced the following expressions for ratio of fission width ($\Gamma_{f}$) to total decay width ($\Gamma_{total}$) for $T_0-\dfrac{1}{2}$, 

\begin{equation}
\dfrac{\Gamma_{f}^{T_0-1/2}}{\Gamma_{total}^{T_0-1/2}}=\dfrac{\langle \sigma_{\alpha,f} \rangle} {\sigma_{\alpha}}
\end{equation}
and $T_0+\dfrac{1}{2}$ states of CN,

\begin{equation}
\dfrac{\Gamma_{f}^{T_0+1/2}}{\Gamma_{total}^{T_0+1/2}}=(2T_0+1) [\dfrac{\langle \sigma_{p,f} \rangle} {\sigma_{p}}-\dfrac{\langle \sigma_{\alpha,f} \rangle} {\sigma_{\alpha}}]
\end{equation}

From Eq. (1) and (2), we can calculate the fission branching ratios for $T_0-\dfrac{1}{2}$ and $T_0+\dfrac{1}{2}$ states, respectively. For $^{209}$Bi(p, f) and  $^{206}$Pb($\alpha$, f) datasets, the value of $T_0= 21.5$. Yadrovsky has taken the experimental data of the ratio of cross-sections $\dfrac{\langle \sigma_{p,f} \rangle} {\sigma_{p}}$ and $\dfrac{\langle \sigma_{\alpha,f} \rangle} {\sigma_{\alpha}}$ for $^{209}$Bi(p, f) and  $^{206}$Pb($\alpha$, f), respectively from Gadioli $\it{et}$ $\it{al.}$~\cite{gadioli}. Yadrovsky, further, calculated the fission widths by considering total decay width to be equal to the sum of neutron decay width, proton decay width and fission width and then, obtained the proton and neutron decay width using optical model. But in our work, we limit ourselves to fission branching ratios for two isospin states of CN by using a different experimental dataset.

\section{Results and Discussion}
\label{sec-2} 
\subsection{$^{209}$Bi(p, f) and  $^{206}$Pb($\alpha$, f)}
\label{sec-3}
First, we look for the experimental data of fission cross-sections for $^{209}$Bi (p,f) and $^{206}$Pb ($\alpha$,f) at different excitation energies of CN, $^{210}$Po. For $^{209}$Bi(p, f), we extract the data of fission cross-sections from Fig. 1 of Prokofiev~\cite{prokofiev}. According to Prokofiev, the method of data normalization done by Gadioli $\it{et}$ $\it{al.}$~\cite{gadioli} is not clear and, therefore, unreliable. Therefore, by using different experimental data sets from EXFOR, Prokofiev analyzed and obtained the data for cross sections for many proton induced reactions at different incident energies as shown in Fig. 1 of Ref.~\cite{prokofiev}. For $^{206}$Pb($\alpha$, f), we use the fission cross-section data from Huizenga $\it{et}$ $\it{al.}$~\cite{huizenga}. Huizenga $\it{et}$ $\it{al.}$~\cite{huizenga} reported experimental X-sections at five different projectile energies. We fit this curve exponentially so that we can have fission X-sections at excitation energies of CN same as obtained for $^{209}$Bi(p, f). We calculate the CN fusion cross-sections $\sigma_{p}$ and $\sigma_{\alpha}$ for $^{209}$Bi (p,f) and $^{206}$Pb ($\alpha$,f) using PACE4~\cite{pace} at different energies of projectiles. After getting all the required quantities, we put these values in Eq. (1) and (2) to calculate the fission widths for $T_0-\dfrac{1}{2}$ and $T_0+\dfrac{1}{2}$ states of CN, table~\ref{tab-1}. In the table~\ref{tab-1}, $E_{p}$ and $E_{\alpha}$ represent the incident energies of projectiles, proton and alpha and $E^{*}_{CN}$ represents the excitation energy of CN.

\begin{figure}[h]
\centering
\includegraphics[width=8cm]{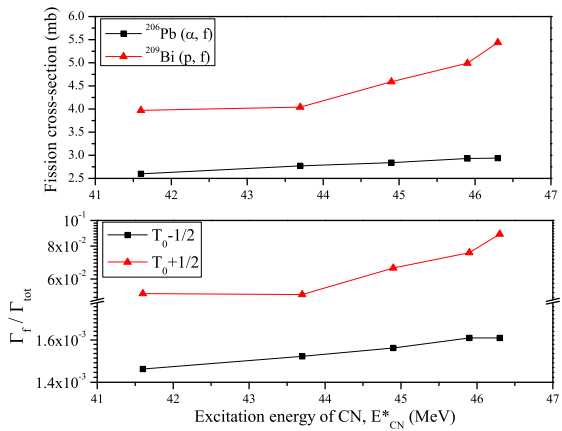}
\caption{Comparison of fission cross sections $\sigma_{p,f}$ and  $\sigma_{\alpha,f}$ for $^{209}$Bi (p,f) and $^{206}$Pb ($\alpha$,f), respectively (upper panel). Comparison of fission branching ratios for two isospin states of CN, $T_0+\dfrac{1}{2}$ and $T_0-\dfrac{1}{2}$ (lower panel).}
\label{firstset}       % Give a unique label
\end{figure}

In Fig.~\ref{firstset}, we compare the fission cross-section for $^{209}$Bi (p,f) and $^{206}$Pb ($\alpha$,f) and fission branching ratios for $T_0+\dfrac{1}{2}$ and $T_0-\dfrac{1}{2}$ states of CN, $^{210}$Po using table~\ref{tab-1}. We can see that there is a reasonable difference between the fission branching ratios for two isospin states of CN. The fission branching ratio for $T_0+\dfrac{1}{2}$ is greater by an order than for $T_0-\dfrac{1}{2}$ state. This supports Yadrovsky's claim that CN remembers the isospin value during fission.

\begin{table*}
\centering
\caption{Ratio of fission width to total decay width for $T_0+\dfrac{1}{2}$ and $T_0-\dfrac{1}{2}$ states of CN at different $E^{*}_{CN}$ by using experimental data of fission cross-section for $^{209}$Bi(p,f) and $^{206}$Pb($\alpha$,f) datsets.}
\label{tab-1}       % Give a unique label
% For LaTeX tables you can use
\begin{tabular}{|l|l|l|l|l|l|l|l|l|}
\hline
\rule{0pt}{20pt} $E_{p}$ & $E_{\alpha}$ & $E^{*}_{CN}$ &  $\sigma_{p,f}$ & $\sigma_{\alpha,f}$ & $\sigma_{p}$ & $\sigma_{\alpha}$ & $\dfrac{\Gamma_{f}^{T_0+1/2}}{\Gamma_{total}^{T_0+1/2}}$ & $\dfrac{\Gamma_{f}^{T_0-1/2}}{\Gamma_{total}^{T_0-1/2}}$ \\(MeV) & (MeV) & (MeV) & (mb) & (mb) & (mb) & (mb) &&\\
\hline
\rule{0pt}{10pt} 36.7 & 47.9 & 41.6 & 3.97 & 2.6 &  $1.49 \times 10^{3}$ & $1.78 \times 10^{3}$ & $5.28 \times 10^{-2}$ & $1.46 \times 10^{-3}$\\
\hline
\rule{0pt}{10pt} 38.9 & 50.1 & 43.7 & 4.04 & 2.77 & $1.49 \times 10^{3}$ & $1.82 \times 10^{3}$ & $5.24 \times 10^{-2}$ & $1.52 \times 10^{-3}$\\
\hline
\rule{0pt}{10pt} 40.1 & 51.3 & 44.9 & 4.59 & 2.84 & $1.5 \times 10^{3}$ & $1.82 \times 10^{3}$ & $6.6 \times 10^{-2}$ & $1.56 \times 10^{-3}$\\
\hline
\rule{0pt}{10pt} 41.1 & 52.3 & 45.9 & 4.99 & 2.93 & $1.5 \times 10^{3}$ & $1.82 \times 10^{3}$ & $7.55 \times 10^{-2}$ & $1.61 \times 10^{-3}$\\
\hline
\rule{0pt}{10pt} 41.5 & 52.7 & 46.3 & 5.44 & 2.94 & $1.5 \times 10^{3}$ & $1.82 \times 10^{3}$ & $8.87 \times 10^{-2}$ & $1.61 \times 10^{-3}$\\
\hline
\end{tabular}
% Or use
%\vspace*{5cm}  % with the correct table height
\end{table*}

\section{Conclusion}
We have calculated the ratio of fission width to total decay width or fission branching ratio for $T_0+\dfrac{1}{2}$ and $T_0-\dfrac{1}{2}$ states of CN at different excitation energies of CN using Yadrovsky's idea. Yadrovsky has done the calculation for a combination of $^{209}$Bi (p,f) and $^{206}$Pb ($\alpha$,f) giving a CN $^{210}$Po and found that there is a noticeable difference of the order of $10^{2}$-$10^{3}$ between fission branching ratios for two isospin states of CN. For fission decay width, this difference goes upto $10^{5}$. We have done the calculation of fission branching ratios for the same combination of datasets taken by Yadrovsky but with different experimental data. We also use PACE4~\cite{pace} to calculate CN formation cross-section at different projectile energies for different sets of reactions. There is reasonable difference of fission branching ratios for two different isospin states of CN. This reinforces the claim made by Yadrovsky that a nucleus remembers isospin of the initial states leading to fission. This also provides further support to our claim of isospin conservation in fission as presented in~\cite{jain, swati}. 

\section*{Acknowledgment}
Support from Ministry of Human Resource Development (Government of India) to SG in the form of a fellowship is gratefully acknowledged. The authors also acknowledge the financial support in the form of travel grant from IIT Roorkee Alumni funds.

\end{document}